\documentclass[twocolumn]{aastex6}

\fullcollaborationName{ Co-authors and affiliations can be found after the acknowledgements}

\begin{document}



\title{Discovery of a Metal-Poor Field Giant with a Globular Cluster Second-Generation Abundance Pattern}


\author{J. G. Fern\'andez-Trincado\altaffilmark{1},
	A. C. Robin\altaffilmark{1},
	E. Moreno\altaffilmark{2},
	R. P. Schiavon\altaffilmark{3},
	A. E. Garc\'ia P\'erez\altaffilmark{4,5},
	K. Vieira\altaffilmark{6},
	K. Cunha\altaffilmark{7},
	O. Zamora\altaffilmark{4, 5},
	C. Sneden\altaffilmark{8},
	Diogo Souto\altaffilmark{9},
	R. Carrera\altaffilmark{4, 5},
	J. A. Johnson\altaffilmark{10},
	M. Shetrone\altaffilmark{11},
	G. Zasowski\altaffilmark{12},	
	D. A. Garc\'ia-Hern\'andez\altaffilmark{4, 5},
	S. R. Majewski\altaffilmark{13},	
	C. Reyl\'e\altaffilmark{1},
	S. Blanco-Cuaresma\altaffilmark{14},
	L. A. Martinez-Medina\altaffilmark{2}, 
	A. P\'erez-Villegas\altaffilmark{15},
	O. Valenzuela\altaffilmark{2},
	B. Pichardo\altaffilmark{2},
	A. Meza\altaffilmark{16},
	Sz.~M{\'e}sz{\'a}ros\altaffilmark{17},
	J. Sobeck\altaffilmark{13},
	D. Geisler\altaffilmark{18},
	F. Anders\altaffilmark{19, 20},
	M. Schultheis\altaffilmark{21},
	B. Tang\altaffilmark{18},
	A. Roman-Lopes\altaffilmark{22},
	R. E. Mennickent\altaffilmark{18},
	K. Pan\altaffilmark{23},
	C. Nitschelm\altaffilmark{24},
	\& F. Allard\altaffilmark{25}\\
	  (Affiliations can be found after the references)
}



\begin{abstract}

We report on detection, from observations obtained with the APOGEE spectroscopic survey, of a metal-poor ([Fe/H] $= -1.3$ dex) field giant star with an extreme Mg-Al abundance ratio ([Mg/Fe] $= -0.31$ dex; [Al/Fe] $= 1.49$ dex). Such low Mg/Al ratios are seen only among the second-generation population of globular clusters, and are not present among Galactic disk field stars. The light element abundances of this star, 2M16011638-1201525, suggest that it could have been born in a globular cluster. We explore several origin scenarios, in particular studying the orbit of the star to check the probability of it being kinematically related to known globular clusters. We performed simple orbital integrations assuming the estimated distance of 2M16011638-1201525 and the available six-dimensional phase-space coordinates of 63 globular clusters, looking for close encounters in the past with a minimum distance approach within the tidal radius of each cluster. We found a very low probability that 2M16011638-1201525 was ejected from most globular clusters; however, we note that the best progenitor candidate to host this star is globular cluster $\omega$ Centauri (NGC 5139). Our dynamical investigation demonstrates that 2M16011638-1201525 reaches a distance $|Z_{max}| < 3 $ kpc from the Galactic plane and a minimum and maximum approach to the Galactic center of $R_{min}<0.62$ kpc and $R_{max}<7.26$ kpc in an eccentric  ($e\sim0.53$) and retrograde orbit. Since the extreme chemical anomaly of 2M16011638-1201525 has also been observed in halo field stars, this object could also be considered a halo contaminant, likely been ejected into the Milky Way disk from the halo. We conclude that, 2M16011638-20152 is also kinematically consistent with the disk but chemically consistent with halo field stars.\\	
\end{abstract}

\keywords{stars: abundances --- stars: Population II --- globular clusters: general --- Galaxy: structure --- Galaxy: formation}



\section{Introduction}
\label{Chapter1}

\begin{figure}
	\begin{center}
		       \includegraphics[width=95mm,height=85mm]{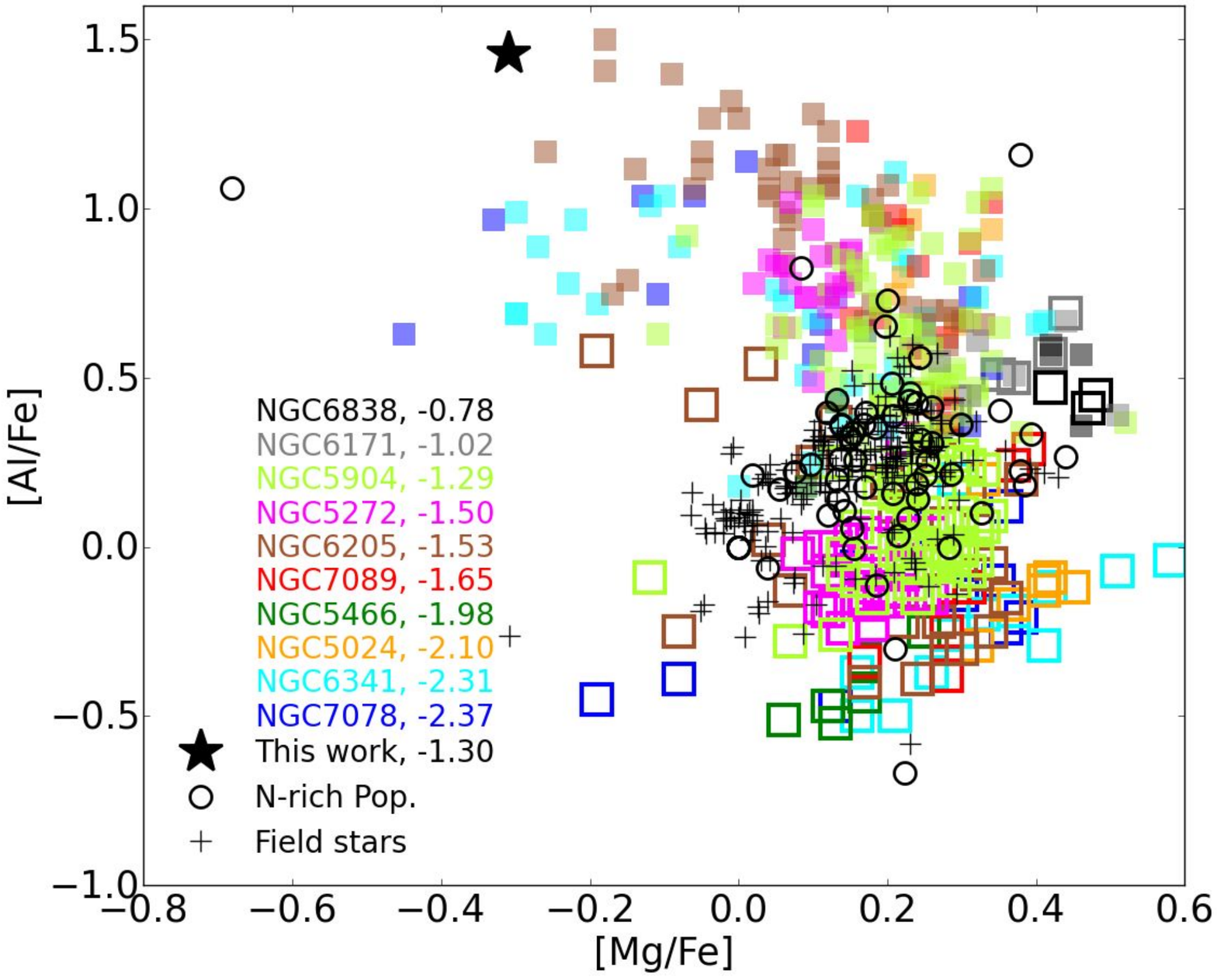}
	\end{center}
	\caption{Al and Mg abundances of 393 red giant stars in 10 GCs from \citet{Meszaros2015}, first generation is marked as open square symbols, and second-generation as filled square symbols. The average metallicity ([Fe/H]) is listed after the cluster name.  The black star symbol represents the star analysed in this work, the black open circles the \citet{Schiavon2016} sample, and plus symbols correspond to 211 stars with available ASPCAP abundances in the same field of 2M16011638-1201525.}
	\label{Figure1}
\end{figure}

It is a commonly accepted observational fact that second-generation stars make up a significant fraction of the population of most Galactic globular clusters \citep[GCs,][]{Carretta2009a, Carretta2009b, Bastian2015},  and they display unique inhomogeneities in their abundance of light elements involved in proton-capture processes. The elements C, N, O, F, Na, Al, Mg, and perhaps Si \citep[e.g.,][and references therein]{Gratton2012, Meszaros2015} provide useful information about the environment in which they were formed. A fraction of GC populations show a pronounced [Al/Fe] and [Mg/Fe] anti-correlation \citep[e.g.,][]{Sneden2004, Johnson2005, Marino2008, Carretta2009a, Carretta2012, Meszaros2015}, a remarkable characteristic (i.e., a \textit{chemical fingerprint}) of some particular second generation stars, { which are not typically observed in the field, except in some peculiar cases, as described below.}

Interestingly, it has been suggested that $\sim$3 \% of field stars in the Galaxy show atypical light-element patterns similar to those seen only among the secondary population 
of Galactic GCs \citep[e.g.,][]{Carretta2010c, Martell2010, Martell2011, Ramirez2012}. They could be explained by the escape of individual stars from those systems \citep[][]{Carretta2010c, Carollo2013,
 Fernandez-Trincado2013,	Carretta20133, 
 Fernandez-Trincado2015a, Fernandez-Trincado2015b, Fernandez-Trincado2016a, Anguiano2016}. 

Only a few studies confirm the existence of such stars in the Galactic field, for example: (\textit{i}) some Aquarius stream stars appear to originate from a globular cluster \citep[e.g.,][]{WyliedeBoer2012}; (\textit{ii}) The \citet{Carretta20133} study on NGC  6752, identifies a few field stars candidates with clear chemical patterns of GCs; (\textit{iii}) \citet[][]{Ramirez2012} found an elevated Na/O ratio abundance in two field halo dwarf stars; (\textit{iv}) \citet[][]{Lind2015} recently discovered a metal-poor field halo star with high Al-Mg ratio; (\textit{v}) { \citet{Martell2016} recently identified five stars in the Galactic halo with GC-like abundance patterns, and these stars are thought to be migrants from GCs}; (\textit{vi}) \citet{Schiavon2016} analysed chemical compositions of 5,175 stars in fields centered on the Galactic bulge, and found 59 giant stars with elevated nitrogen abundances, anti-correlated with [C/Fe] and correlated with [Al/Fe] abundance; several scenarios have been put forward to explain such anomalies in the Galactic bulge, i.e., the authors proposed that these stars may likely come from disrupting GCs, though alternative scenarios being considered that N-rich stars could be formed in environments similar to the GCs \citep[for more details, see][]{Schiavon2016}. Detection of such light-element abundance inhomogeneities in non-GC stars, are particularly important for understanding how many Galactic field stars could have been deposited by GCs.

In this work, we report the first discovery in APOGEE data of a peculiar giant star in the Milky Way field that stands out in its very low  magnesium and high aluminium abundance, as well as other very significant light-element abundance anomalies, such as a greatly enhanced nitrogen to iron ratio ([N/Fe]$>$1.0). This star could be the most convincing evidence yet for a Galactic field star stripped from a GC.

\section{A peculiar giant star observed by APOGEE}
\label{Chapter2}

The star of interest,  2M16011638-1201525, was found in the Apache Point Observatory Galactic Evolution Experiment \citep[APOGEE, ][]{Zasowski2013, Majewski2015}, a near-infrared spectroscopic survey, part of the Sloan Digital Sky Survey III \citep[SDSS-III;][]{Eisenstein2011}, targeting primarily Milky Way red giants, at a resolution of R$\approx$22,500, acquired with the APOGEE multi-object spectrograph mounted at the SDSS 2.5m telescope \citep{Gunn2006}.  We refer the reader to \citet{Holtzman2015} and \citet{Nidever2015} for detailed information on the data, and the data reduction pipeline. 

We turn our attention to the giant star 2M16011638-1201525, with high [Al/Fe]$=1.49$ and a strongly depleted [Mg/Fe]$=-0.31$ (manually confirmed using MOOG\footnote{\url{http://www.as.utexas.edu/~chris/moog.html}}). 2M16011638-1201525 has been identified as part of a sample of $\sim 265$ giant stars originally surveyed spectrocopically by APOGEE in a field centered on (\textit{l,b}) = (0, +30), having unusual chemical abundances, and a high quality stellar radial velocity. Figure \ref{Figure1} shows a comparison of our results with respect to those derived from 393 giants in 10 GC stars \citep[see][]{Meszaros2015} and 59 bona fide field nitrogen rich stars giants in the bulge from \citet[][]{Schiavon2016}, in particular we show that 2M16011638-1201525 has one of the most extreme combinations of abundances. 2M16011638-1201525 shows a radial velocity (with typical uncertainties of less than 1 km s$^{-1}$) dispersion ($vscatter$) less than 1 km s$^{-1}$ over 4 visits,  which makes it unlikely to be a variable star or a binary star. We also examine variations between 2MASS and DENIS magnitudes and USNO-B catalogs, and find no evidence for photometric variations between those catalogs, i.e., $(K_{2MASS}-K_{DENIS})=-0.016$ mag, and $(R1-R2)=0.07$ mag. 

Figure \ref{Figure1} shows the light-element anomalies of 2M16011638-1201525,  in this case for Al and Mg abundances. Such extreme values of Al enhancements and Mg depletions are only observed in second-generation GC population, as seen in Figure \ref{Figure1}.

The main atmopheric parameters (T$_{\rm eff}$, log \textit{g}, and [Fe/H]) of  2M16011638-1201525 were checked using an extended and updated version of iSpec\footnote{\url{http://www.blancocuaresma.com/s/iSpec/}} \citep{BlancoCuaresma2014} to work in the spectral regime of APOGEE ($\sim$1.51 $\mu{}m$ to 1.7$\mu{}m$). For a set of atmospheric parameters, and atomic data, iSpec generates synthetic spectra, computed from the ATLAS atmosphere model, \citet{Kurucz2005}, and minimises the difference with the observed spectrum using a least-squares algorithm. 

We adopt the iSpec recommended stellar parameters: T$_{\rm eff} = 4572\pm100$ K, log \textit{g}$=1.66\pm0.1$, and [Fe/H]$=-1.30\pm0.1$), which are entirely  consistent with  those obtained by ASPCAP pipeline \citep{Garcia2015}, T$_{\rm eff} = 4575\pm92$ K, log \textit{g}$=1.61\pm0.11$, and [Fe/H]$=-1.31\pm0.05$. Both sets of model parameters computed with the ATLAS model grid are consistent with those we found using MARCS stellar atmospheres models \citep{Gustafsson2008, Zamora2015}.

 In this work\footnote{To facilitate the reproducibility and reuse of our results, we have made all the simulations available in a public repository at \url{https://github.com/Fernandez-Trincado/SDSS-IV-Project0184/blob/master/README.md}}, we focus on the abundances of Al, Mg, C, N, and O, which are typical chemical signatures of GCs \citep[][]{Gratton2012}. We did not include sodium in our analysis, which is a typical species to separate GC's populations, as its lines in our APOGEE spectra (1.6373$\mu$m and 1.6388$\mu$m) are weak in the typical T$_{\rm eff}$ and metallicity for the star studied in this work, and this would lead to unreliable abundance results.
	
\begin{figure*}
	\begin{center}
		\includegraphics[width=1.0\textwidth]{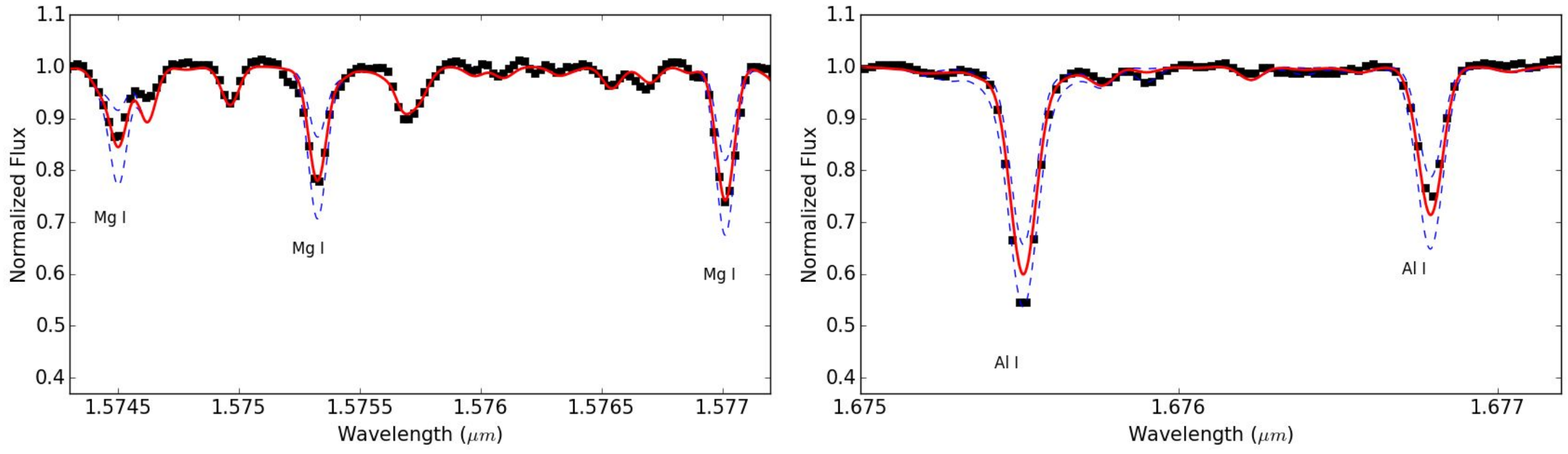}
	\end{center}
	\caption{The best-fit for the Mg {\sc i} and Al {\sc i}  lines (red curve) in the observed infrared spectrum (black filled squares) of 2M16011638-1201525 with S/N $>$700. The two blue dashed curves correspond to synthetic spectrum abundance choices that are offset from the best fit by $\pm$0.5 dex.}
	\label{Figure2}
\end{figure*}

	APOGEE spectra has three main windows to determine aluminium abundances: 1.6718$\mu$m, 1.6750$\mu$m, and 1.6763$\mu$m. 
	We did not analyze the line at 1.6718$\mu$m because it is poorly fitted in the core and this may be an indicator of NLTE or saturation effects \citep[][]{Hawkins2016}. The selected lines at 1.6750$\mu$m and 1.6763$\mu$m shown an offset of $\pm0.5$ dex (see Fig. \ref{Figure2}) between the two best-fit abundances, i.e., the derived line-to-line abundance is A(1.6750$\mu$m) = 6.81 and A(1.6763$\mu$m) = 6.31. It is important to note that this discrepancy does not affect the discusion and conclusions of this work, i.e., the line-to-line and relative abundance indicates that the star is Al-rich. For 2M16011638-1201525 we have done a manual inspection of the best MOOG \citep[v. Jan2016, ][]{Sneden1973} fitted synthesis of Al, Mg, C, N, and O lines, using atomic and molecular species, the most recent OH line list by \citet{Brooke2016}, and Solar abundances values from \citet{Asplund2005}. The best MOOG fitted synthesis of Al and Mg lines for 2M16011638-1201525 are shown in Figure \ref{Figure2}. For our manual analysis we adopted the best fit of atmospheric stellar parameters recommended by iSpec in good agreement with other independent analyses and methods. This step was necessary to provide a consistent comparison of the results from a manual abundance analysis with the values determined from the ASPCAP pipeline. Table \ref{table1} gives these abundances, and those as derived by ASPCAP pipeline using a different line list of atomic and molecular species. For comparison the abundances derived from the photometric effective temperatures is given in the same table. 

We additionally computed abundances assuming effective temperature from photometry to check for any significant deviation in our results, i.e, a photome\-tric effective temperature was calculated from the $J-K$ colors relation using the methodology presented in \citet{Gonzalez2009}. Photometry is extinction-corrected using the Rayleigh Jeans Color Excess (RJCE) method \citep[see ][]{Majewski2011}, which leads to an extinction value ${\rm \langle A_K^{WISE} \rangle} \sim 0.157$ mag. For comparison, the Table \ref{table1} shows the values obtained in each procedure. The observed small discrepancies do not affect the main result of our work about the extreme abundances of Mg, Al, and N. 
		
Furthermore, our abundances were compared with the literature, i.e., chemical abundances from the DR12 data \citep[][]{Garcia2015}, and GC's stars \citep{Meszaros2015}, see Figure \ref{Figure1}. Their values are quite different from ours, as also seen in the online\footnote{DR12 Science Archive Server (SAS):  \url{http://dr12.sdss3.org/irSpectrumDetail?commiss=0&locid=4520&show_aspcap=True&apogeeid=2M16011638-1201525}} version from the best ASPCAP fit. This is due to the fact that we used the same wavelength windows than \citet[][see their Table 3]{Meszaros2015}, which are significantly different from the ASPCAP DR12 windows. Also we did not use any of the weak Mg lines, which in these metal poor stars mostly disappear from the spectra. These differences in [Mg/Fe], [Al/Fe], [N/Fe], [C/Fe], [N/Fe], and [O/Fe] abundances are also likely to be due to the updated line list which includes both atomic and molecular species used by MOOG in our procedure. The chemical abundances relevant to this work are not affected by the analysis methods used.

\begin{table}
	\setlength{\tabcolsep}{2.5mm}  
	\begin{tiny}
		\caption{Chemical abundances of 2M16011638-1201525}
		\label{table1}
		\begin{tabular}{ccccc}
			\hline
			\hline
			&APOGEE DR12               &   This work                         &  Photometric \\ 
				&  T$_{\rm eff} $   &   T$_{\rm eff} $  &   T$_{\rm eff} $ \\
		&  $ 4575$  K      K &  $ 4572$  K   &   $4340$ K  \\
			\hline
			\hline
		$	{\rm [Fe/H]} $   & -1.31  &  -1.30    &    -1.30   \\
		$	{\rm [C/Fe]} $   & 0.09	    &   -0.15   &     -0.06    \\
		$	{\rm [N/Fe]} $   & 1.08	    &     1.46    &     1.05  \\
		$	{\rm [O/Fe]}$    & 0.21     &     -0.06  &     0.03  \\
		$	{\rm [Mg/Fe]} $ &  -0.04  &     -0.31  &   -0.43  \\
		$	{\rm [Al/Fe]} $   &  1.06   &   1.49    &     1.28   \\
			\hline
			\hline
		\end{tabular} 
			\tablecomments{The Solar reference abundances are from \citet{Asplund2005}.}
	\end{tiny}
\end{table}

\begin{table}
	\setlength{\tabcolsep}{4.0mm}  
	\begin{tiny}
		\caption{Phase-space data}
		\label{table2}
		\begin{tabular}{ll}
			\hline
			\hline

	Coordinates & (J2000)     \\
	\,\,\,$(\alpha,\, \delta)$	& $(240^\circ .31825,\, -12^\circ .03127)$ \\
	\,\,\,$(l,\, b)$	& $(358^\circ .87794,\, 29^\circ .5692)$\\  
	Heliocentric Distance    &  [kpc]  \\
	                                      & $(2.67\pm 0.68)^{\rm a}$   \\
	                                      & $(2.94\pm 0.62)^{\rm b}$    \\
	$V_{los}$ & ${\rm \,[km\, s^{-1}}]$\\
	                                   &(82.23$\pm$0.84)$^{\rm a}$  \\
	                                  & (84.68$\pm$0.79)$^{\rm c}$ \\
	Proper Motions   & $(\mu_\alpha\cos\delta\, ,\,\, \mu_\delta)$ \\
                      	      & (${\rm mas\,yr^{-1}}$) \\
	                          & $(-11.5\pm1.7,-16.9\pm1.7)^{\rm d}$ \\
	                          & $(-12.3\pm2.1,-16.0\pm2.1)^{\rm e}$ \\
	                          & $(-15.7\pm2.7,-17.2\pm2.4)^{\rm f}$ \\

			\hline
			\hline
		\end{tabular} 
		\tablecomments{$^{\rm a}$\citet{Kordopatis2013a}; 
	  $^{\rm b}$\citet{Hayden2014}; $^{\rm c}$SDSS-III/APOGEE; $^{\rm d}$UCAC4; $^{\rm e}$PPMXL; $^{\rm f}$Tycho-2 ( unfortunately improvements in distance and proper motions are not available from TGAS \citep{Gaia2016} catalogue for this star).}
	\end{tiny}
\end{table}

\section{Galactic model and simulations}
\label{models}

We performed a series of orbital\footnote{We have adopted a righthanded coordinate system for (U, V, W), so that they are positive in the directions of the galactic center, galactic rotation, and north galactic pole, respectively.}  integrations using a semi-analytical,
multicomponent model of the Milky Way potential to predict the orbital
parameters of 2M16011638-1201525 in the Galaxy, based on the 
reliable (Table \ref{table2}) six-dimensional phase space coordinates (3D position and
3D velocity).

We consider axisymmetric and non-axisymmetric Galactic
models including a prolate bar and spiral arms structures.  The relevant parameters employed in the
bar and the spiral arms are the same as those explained in \citet[][]{Moreno2014, Robin2012}.

We employed the kinematical parameters of 2M16011638-1201525 and those associated with the Galactic model, and consider their corresponding uncertainties
as 1$\sigma$ variations in a Gaussian Monte Carlo sampling.
The adopted Solar motion with its uncertainties is 
$(U,V,W)_{\odot}$= ($-11.1\pm$1.2, 12.24$\pm$2.1, 7.25$\pm$0.6) 
km s$^{-1}$ \citep[e.g.,][]{Schonrich2010, Brunthaler2011}. In each computed orbit we obtain the following orbital
parameters: maximum distance from the Galactic plane, $Z_{max}$;
maximum and minimum Galactocentric radii, $r_{max}$ and $r_{min}$; 
and the orbital eccentricity defined as
$e =(r_{max}-r_{min})/(r_{max}+r_{min})$.

To estimate the effect of axisymmetric and non-axisymmetric components 
of the Galactic potential in the computed orbital parameters, we
considered the following four configurations of the Galactic potential:

\begin{enumerate}

	\item[(i)] { Model 1:} the axisymmetric model, which is the direct scaling of the \citet{Allen1991} model.
	
	\item[(ii)] { Model 2:} the non-axisymmetric model
mentioned above, using the prolate bar with the spiral arms \citep[see][]{Pichardo2003, Pichardo2004}. 
  
    \item[(ii)] { Model 3:} the non-axisymmetric model using the boxy bar with the spiral arms \citep[see][]{Pichardo2003, Pichardo2004}. 
    
     \item[(iv)] { Model 4:} We also performed orbit integration of 2M16011638-1201525 using the  \textit{GravPot16}\footnote{\url{https://fernandez-trincado.github.io/GravPot16/}} code. We have assumed a Milky Way's gravitational potential model based on the mass distribution of the last version of the Besan\c{c}on Galaxy model \citep[see ][]{Robin2014, Fernandez-Trincado2014} based on the superposition of many composite stellar populations belonging to the thin disk, dark matter halo component, and interstellar matter \citep[e.g.,][]{Robin2003}, the thick disk \citep[shape B, see ][]{Robin2014} and stellar halo \citep[][]{Robin2014, Fernandez-Trincado20152}, and a standard triaxial boxy shape bar \citep{Robin2012}. The results are shown in Figure \ref{Figure4}. We refer the reader to these papers for further details of the density profiles.
\end{enumerate}
    
In these models the orbits of the star are integrated over
2 Gyr, with $10^4$ orbits in Model 1 and $10^3$ orbits in both, Model 2 and
Model 3. We have found that in Models 2, 3  and 4 the computed orbital parameters
of 2M16011638-1201525 are very similar, thus not depending sensitively on
the assumed model of the Galactic bar.

 For the orbital computation of 2M16011638-1201525, we also tested the set of distances and proper motions given in Table \ref{table2}; small variations in these observables do not lead to substantial difference in the orbital parameters and does not affect the overall conclusions of this work. 
  
 We chose to use the accurate ($< 1$ km s$^{-1}$) radial velocity from APOGEE. We adopted the spectro-photometric estimated distance from \citet{Hayden2014} based on Bayesian methods developed for APOGEE data, which is in good agreement with the distance measurements from RAVE survey (see Table \ref{table2}). We have adopted absolute proper motions from the UCAC4 catalogue \citep[Fourth U.S. Naval Observatory CCD Astrograph Catalogue][]{Zacharias2013}, because the error in proper motion  (with uncertainties $<$ 2 mas yr$^{-1}$) is smaller compared to other catalogues (see Table \ref{table2}), and is less affected by potential systematic uncertainties \citep{Vickers2016}. We note that the small uncertainties on the proper motions are good enough to estimate the space-velocity vector accurately, i.e.,  ($U_{\rm LSR}$, $V_{\rm LSR}$, $W_{\rm LSR}$)\footnote{The velocities ($U_{\rm LSR}, V_{\rm LSR}, W_{\rm LSR}$) are relative to the local standar of rest (LSR)} = (93.9$\pm$10.9, -246.1$\pm$68.9, 21.7$\pm$19.9) km s$^{-1}$.
  
\section{POSSIBLE ORIGINS}
\label{Discussion}

A possible scenario producing very extreme Mg-Al-N nucleosynthesis could be an association with an intermediate mass ($\sim 3 - 6$ M$_{\odot}$) AGB star \citep[see][]{Ventura2011, Schiavon2016} in a binary companion. A future work will be dedicated to investigate in more detail other mechanisms, including binary stellar mergers or pollution of the interstellar matter (ISM) by a previous generation of massive stars. 

In the following subsections we will analyze other scenarios that could have led this peculiar star to its current phase space location in the Galaxy. 

\subsection{A dwarf spheroidal  (dSph) galaxy interloper?}

\citet[][]{Lind2015} argue that stars with low [Mg/Fe]-ratios are commonly found in dwarf spheroidal galaxies. However, [Al/Fe] enhancement is not expected in these systems \citep[see ][]{Koch2008}. Therefore, given the high Al enhancement and strongly depleted Mg observed in 2M16011638-1201525, { we conclude that it seems unlikely that a merged and disrupted dwarf spheroidal galaxy could have hosted 2M16011638-1201525.}

\subsection{A globular cluster escapee?} 
 
In order to study the ejection scenario of 2M16011638-1201525 from a not entirely disrupted GC into the Milky Way disk, we performed a kinematical analysis using Model 2 over a 2 Gyrs period. It is based on $3\times10^3$ Monte Carlo cluster and 2M16011638-1201525 orbits for a sample of 63 Galactic GC with good proper motion measurements \citep[see ][]{Moreno2014}. 

\begin{figure}
	\begin{center}
		\includegraphics[width=0.50\textwidth]{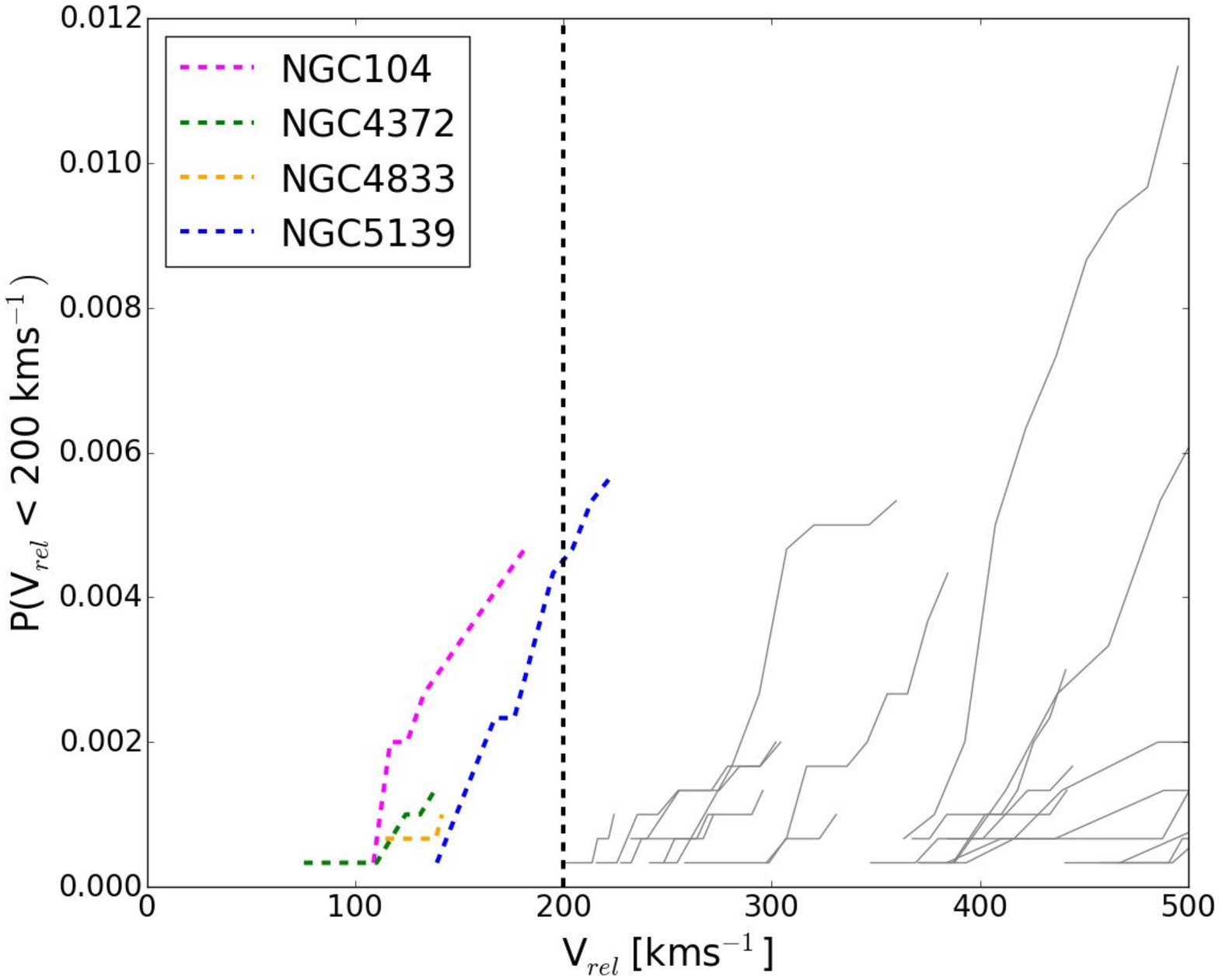}
	\end{center}
	\caption{Cumulative probability distribution of the relative velocities ($V_{rel}$) during close enconters between 2M16011638-1201525 and Galactics GC from \citet{Moreno2014}. The vertical dashed line shows the velocity threshold (200 km s$^{-1}$) adopted in this work. }
	\label{FigureA}
\end{figure}

We assumed that 2M16011638-1201525 could have been ejected from a given GC, with a relative velocity below a certain threshold  $( V_{rel}< 200$ km s$^{-1})$, which may be possible in the interaction of black holes and/or binary systems \citep[see ][]{Fernandez-Trincado2015a}. Then, we computed the cumulative probability distribution (see Figure \ref{FigureA}) for the relative velocity ($V_{rel}$), which is defined as the relative velocity during each close encounter, occurring at times $t<2$ Gyr and within a distance less than or equal to the tidal radius of the GC \citep[$\delta r < r_t$; ][]{Moreno2014}. If a close past encounter is probable, then the GC could be identified as a possible progenitor of 2M16011638-1201525.

Our results suggest that our hypothesis of 2M16011638-1201525 being ejected from a given GC is negligible for 59 of them, and very small ($< 0.5 \%$) for four GC in our sample (see Figure \ref{Figure2}): NGC 5139 ($\omega$ Cen), NGC 104 ($47$ Tucanae), NGC 4372, NGC 4833.

Other parameters also disfavor the proposes hypothesis: 2M16011638-1201525 is metal-poor, [Fe/H]$=-1.30$ dex, with orbital parameters that reach a distance  $|Z_{max}| \sim 3 $ kpc from the Galactic plane (see Figure \ref{Figure4}), which are not consistent with the metallicity \citep[][]{Harris1996} and orbital properties \citep[][]{Moreno2014} for three of the four clusters mentioned above: NGC 104  ([Fe/H]$=-0.72$ dex, $|Z_{max}|\sim$ 3.13 kpc),  NGC 4372 ([Fe/H]$=-2.17$ dex, $|Z_{max}|\sim$1.57 kpc) and NGC 4833 ([Fe/H]$=-1.85$ dex, $|Z_{max}|\sim$1.54 kpc). Both NGC 4372 and NGC 4833, are more metal-poor than 2M16011638-1201525, while NGC 104 is more metal-rich and particularly with a more Mg-rich population, \citep[see][]{Carretta2013} than 2M16011638-1201525, which makes these 3 clusters an unlikely origin for the star. 

Our chemodynamical results show that among the more probable GCs candidates associated with 2M16011638-1201525, NGC 5139 could be a possible progenitor system. This cluster is metal-poor, [Fe/H]$=-1.53$ dex \citep{Harris1996}, and has an orbit that reaches a distance  $|Z_{max}|\sim$1.69 kpc from the Galactic plane \citep{Moreno2014}. Additionally, NGC 5139 shows enrichement levels spanning from $\sim -1.8$ dex to $ \sim-0.5$  dex
\citep[e.g.,][and references therein]{Villanova2014}, and atypical light-element abundances 
with a pronounced Mg-Al anti-correlation \citep[e.g.,][]{Norris1995} like that seen in 2M16011638-1201525. From Figure \ref{FigureA}, we conclude  that this star could have been radially ejected in any direction from NGC 5139
(blue dashed line in Figure \ref{FigureA}). 2M16011638-1201525 has a velocity greater than the cluster's escape velocity, i.e., $V_{rel}> 60.4$ km s$^{-1}$ \citep[see][and references therein]{Fernandez-Trincado2015a}, reaching a total energy (\textit{E}) and angular momentum (\textit{Lz}) from the ejection process 
slightly similar to that of NGC 5139. We note that 2M16011638-1201525 is moving on a retrograde orbit, and have specific angular momenta, \textit{Lz}$=-307$ km s$^{-1}$kpc, similar to that of NGC 5139, i.e., \textit{Lz}$=-342.5$ km s$^{-1}$kpc \citep[see ][]{Moreno2014}; this result could strengthen the association of this star with NGC 5139. 

We emphasize that there is evidence which suggests NGC 5139 as a dominant contributor of retrograde stars, and of stars with chemical anomalies generally found only within GCs \citep{Altmann2005, Majewski2012, Fernandez-Trincado2015b}. These stellar debris and the newly discovered star strongly suggest that NGC 5139 was not formed on its presente orbit, and has been affected by frequent passages through the disk \citep[e.g.,][]{MEZA2005}. Hence many of the claimed stellar debris to be part of NGC 5139 follow orbital properties ($R_{min}, R_{max}, Z_{max}$) slightly different from the host system.

We also highlight that 2M16011638-1201525 has halo-like radial velocity based on the kinematics predicted by the revised version of the Besan\c{c}on Galaxy model \citep{Robin2014}, implying that this star could be interpreted also as a ''halo interloper" especially given a retrograde motion, and a peculiar chemical fingerprint that is consistent with the "GC" halo population.

\subsection{A Galactic bulge interloper?} 

\citet{Schiavon2016} have recently discovered a new stellar population in the Galactic bulge (called the N-rich population), which clearly shows atypical light-element patterns, particularly with elevated nitrogen abundance [N/Fe] $ > $ 1.0 dex. Such abundances are very different from what is seen in the normal stellar population 
of the Galactic bulge in the same spatial region (i.e., $|b|<16^{\circ}$, $-20^{\circ} < l <20^{\circ}$, and 5 kpc $ < d_{\odot} < 11$ kpc). 

\begin{figure*}
	\begin{center}
		\includegraphics[width=0.54\textwidth]{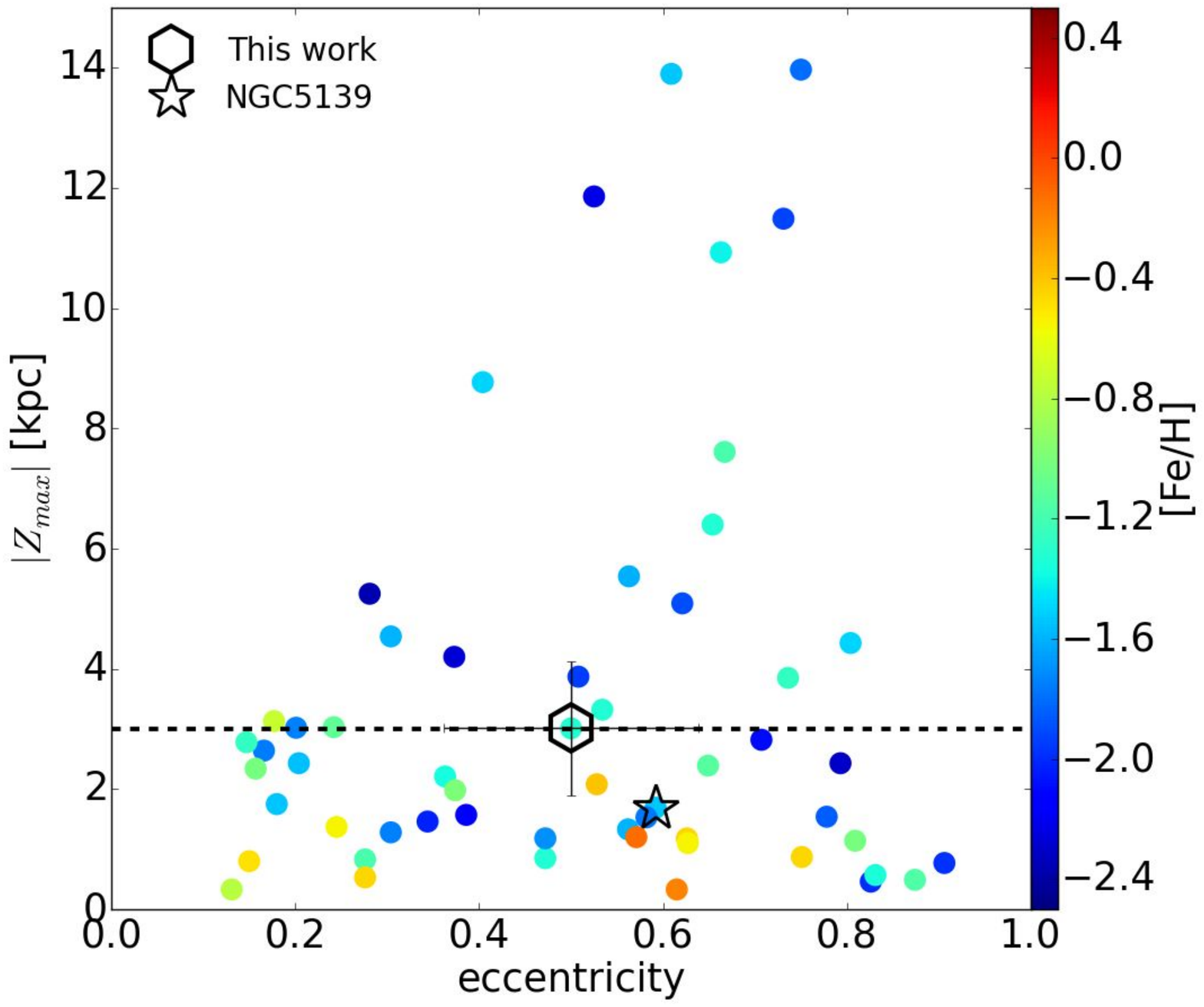}\includegraphics[width=0.42\textwidth]{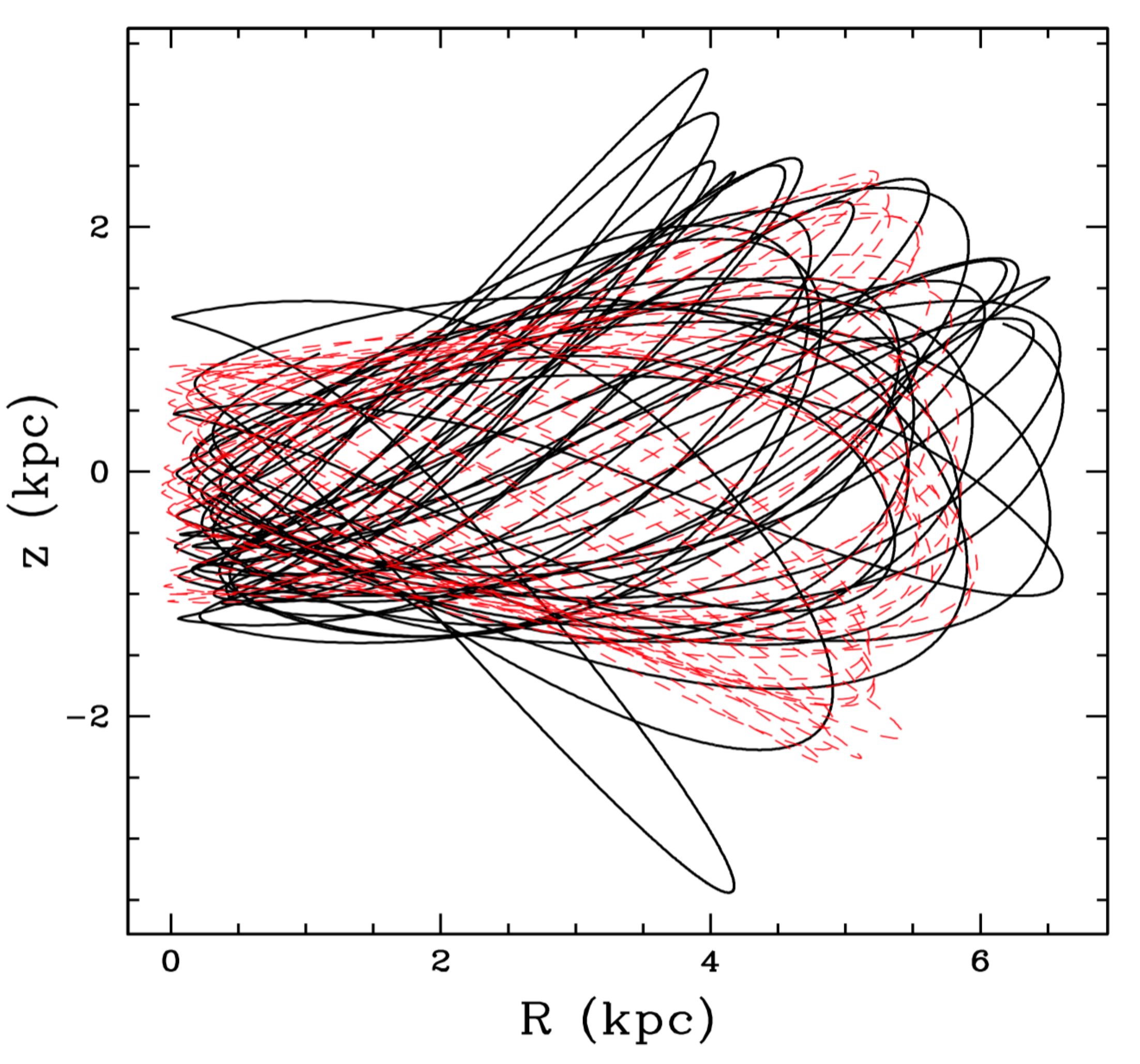}
	\end{center}
	\caption{Left: Maximum distance $|Z_{max}|$ from the Galactic plane as function of the orbital eccentricity for 2M16011638-1201525, and 63 GC from \citet{Moreno2014}. The open star symbol refers to NGC 5139. The horizontal black dashed line represent 3 kpc, the higher limit for the thick disc proposed by \citep{Carollo2010}. Right: Meridional orbit of 2M16011638-1201525 computed with the Model 3 (black line), and Model 4 (red dashed line).}
	\label{Figure4}
\end{figure*}

We consider a possible scenario where 2M16011638-1201525 could be associated with this new stellar population, since this star shows similar nitrogen enhancement as those of the Schiavon's sample, i.e., 2M16011638-1201525 has elevated nitrogen abundance, [N/Fe]$=$1.46 dex (see Table \ref{table1}), like the N-rich population. Interestingly, our orbital solutions show that  2M16011638-1201525  passes through the Galactic bulge at its closest approach to the Galactic center  $R_{min}\sim0.62$ kpc, and reaches a maximum distance from the Galactic centre at $R_{max}<7.26$ kpc (see orbit projection in Figure \ref{Figure4}), in an eccentric orbit $e = 0.53$. Given its peculiar chemical fingerprint and orbital elements within the Galactic disk, this star could be interpreted as a N-rich bulge interloper. { It is interesting to note that there are a handful of N-rich stars from Schiavon's sample with intermediate Al abundance and Mg enhancement (see fig. \ref{Figure1}) making it difficult to link chemically this population with 2M16011638-1201525 which fall outside the main group of field stars, and within the locus of second-generation GC population. However, there is also one star from Schiavon's sample that has Al enhancement and a strongly depleted Mg abundance and is likely within the Al-Mg tail of the N-rich population. On the other hand, one might expect N-rich contaminants with extreme Al-Mg abundance ratio. However, given the kinematic and chemical properties of 2M16011638-1201525, a GC-like second-generation seems more probable.} 

\section{CONCLUSIONS}

We made use of high resolution, near-IR spectra from the SDSS-III/APOGEE survey, and we have discovered the existence of a star within the Milky Way disk with light-element anomalies associated with one of the most extreme combination of Mg and Al anti-correlation seen only in the second generation GC populations. Our orbital computations based on reliable six-dimensional phase space coordinates of this peculiar giant star, 2M16011638-1201525, show that it travels through the Milky Way in a coplanar, eccentric orbit relatively close to the Galactic plane, which suggests that this star has been dynamically ejected into the Milky Way disk from the halo.

A more exotic explanation of such peculiar chemistry in a disk-like orbit star
is that it could be chemically linked with the $\omega$ Cen progenitor system, from which it might have been ejected. However, $\omega$ Cen is a very complex and unusual stellar system in the Milky Way and its origin is still not well understood (globular cluster or dwarf spheroidal galaxy?). Other GC progenitor candidates might be examined with more detail in the near future, given the upcoming and more accurate six-dimensional phase-space data set that will be produced by the Gaia space mission. 

\section*{Acknowledgements}

The authors wish to thank the referee for constructive comments that significantly improved the presentation of this paper.

 We thank UNAM-PAPIIT grants IN 105916 and  IN 114114. J.G.F-T is currently supported by Centre National d'Etudes Spatiales (CNES) through PhD grant 0101973 and the R\'egion de Franche-Comt\'e and by the French Programme National de Cosmologie et Galaxies (PNCG). \\
E.M, L.A.M.M, B.P, and A.P.V acknowledge support from UNAM/PAPIIT grant IN105916.
D.A.G.H was funded by the Ram\'on y Cajal fellowship number RYC-2013-14182. \\
D.A.G.H and O.Z.  acknowledge support provided by the Spanish Ministry of Economy and Competitiveness (MINECO) under grant AYA-2014-58082-P. \\
S.R.M. and C. S. acknowledge support from NSF grants AST-1312863 and AST-1211585, respectively. \\
A.M. acknowledges support from Proyecto Interno UNAB  DI-677-15/N.\\
D.G., and R.M. gratefully acknowledges support from the Chilean  BASAL Centro de Excelencia en Astrof\'isica y Tecnolog\'ias Afines (CATA) grant PFB-06/2007. \\
Szabolcs M{\'e}sz{\'a}ros has been supported by the J{\'a}nos Bolyai 
Research Scholarship of the Hungarian Academy of Sciences.

This research made use of \textit{GravPot16} (Fortran version), a community-developed core under the git version-control system on GitHub. Monte Carlo simulations were executed on computers from the Instituto de Astronom\'ia-UNAM, M\'exico. 

Funding for SDSS-III has been provided by the Alfred P. Sloan Foundation, the Participating Institutions, the National Science Foundation, and the U.S. Department of Energy Office of Science. The SDSS-III web site is http://www.sdss3.org/.

SDSS-III is managed by the Astrophysical Research Consortium for the Participating Institutions of the SDSS-III Collaboration including the University of Arizona, the Brazilian Participation Group, Brookhaven National Laboratory, Carnegie Mellon University, University of Florida, the French Participation Group, the German Participation Group, Harvard University, the Instituto de Astrofisica de Canarias, the Michigan State/Notre Dame/JINA Participation Group, Johns Hopkins University, Lawrence Berkeley National Laboratory, Max Planck Institute for Astrophysics, Max Planck Institute for Extraterrestrial Physics, New Mexico State University, New York University, Ohio State University, Pennsylvania State University, University of Portsmouth, Princeton University, the Spanish Participation Group, University of Tokyo, University of Utah, Vanderbilt University, University of Virginia, University of Washington, and Yale University.


\bibliographystyle{apj}{}
\bibliography{references}



\noindent \hrulefill

\noindent 
${^1}$ Institut Utinam, CNRS UMR 6213, Universit\'e de Franche-Comt\'e, OSU THETA Franche-Comt\'e-Bourgogne, Observatoire de Besan\c{c}on, BP 1615, 25010 Besan\c{c}on Cedex, France; Contact e-mail:
	\href{mailto:jfernandez@obs-besancon.fr}{jfernandez@obs-besancon.fr and/or jfernandezt87@gmail.com}\\
${^2}$ Instituto de Astronom\'ia, Universidad Nacional Aut\'onoma de M\'exico, Apdo. Postal 70264, M\'exico D.F., 04510, Mexico\\
${^3}$ Astrophysics Research Institute, Liverpool John Moores University, 146 Brownlow Hill, Liverpool, L3 5RF, United Kingdom\\
${^4}$ Instituto de Astrof\'{\i}sica de Canarias, 38205 La Laguna, Tenerife, Spain\\
${^5}$ Departamento de Astrof\'{\i}sica, Universidad de La Laguna, 38206 La Laguna, Tenerife, Spain
${^6}$ Centro de Investigaciones de Astronom\'ia, AP 264, M\'erida 5101-A, Venezuela\\
${^7}$ Dept. of Physics and Astronomy, Johns Hopkins University, Baltimore, MD 21210, USA\\
${^8}$ Department of Astronomy and McDonald Observatory, The University of Texas, Austin, TX 78712, USA\\
$^{9}$Observat\'orio Nacional, Rua General Jos\'e Cristino, 77, 20921-400 S\~ao Crist\'ov\~ao, Rio de Janeiro, RJ, Brazil\\
$^{10}$ Department of Astronomy, The Ohio State University, Columbus, OH 43210, USA\\
$^{11}$ Department of Astronomy, University of Texas at Austin, Austin, TX 78712, USA\\
$^{12}$ Department of Physics and Astronomy, Johns Hopkins University, Baltimore, MD 21218, USA\\
$^{13}$ Department of Astronomy, University of Virginia, P.O. Box 400325, Charlottesville, VA 22904-4325, USA\\
$^{14}$ Observatoire de Gen\`eve, Universit\'e de Gen\`eve, CH-1290 Versoix, Switzerland\\
$^{15}$ Max-Planck-Instit\"ut f\"ur Extraterrestrische Physik, Gie\ss enbachstra\ss e, 85748 Garching, Germany\\
$^{16}$ Departamento de Ciencias Fisicas, Universidad Andres Bello, Sazie 2212, Santiago, Chile\\
$^{17}$ ELTE Gothard Astrophysical Observatory, H-9704	Szombathely, Szent Imre Herceg st. 112, Hungary\\
$^{18}$ Departamento de Astronom\'ia, Universidad de Concepci\'on, Casilla 160-C, Concepci\'on, Chile\\
$^{19}$ Leibniz-Institut f\"{u}r Astrophysik Potsdam (AIP), An der Sternwarte 16, 14482, Potsdam, Germany\\
$^{20}$ Laborat\'{o}rio Interinstitucional de e-Astronomia - LIneA, Rua Gal. Jos\'{e} Cristino 77, Rio de Janeiro, RJ - 20921-400, Brazil\\
$^{21}$ Laboratoire Lagrange (UMR7293), Universit\'{e} de Nice Sophia Antipolis, CNRS, Observatoire de la C\^{o}te d'Azur, BP 4229, 06304 Nice Cedex 4, France\\
$^{22}$ Departamento de F\'isica y Astronom\'ia, Universidad de la Serena, Av. Juan Cisternas 1200 Norte, La Serena, Chile\\
$^{23}$ Apache Point Observatory and New Mexico State University, P.O. Box 59, Sunspot, NM, 88349-0059, USA\\
$^{24}$ Unidad de Astronom\'ia, Facultad de Ciencias B\'asicas, Universidad de Antofagasta, 601 Avenida Angamos, Antofagasta, Chile\\
$^{25}$ Univ Lyon, Ens de Lyon, Univ Lyon1, CNRS, Centre de Recherche Astrophysique de Lyon UMR5574, F-69007, Lyon, France.\\

\end{document}